\documentclass[11pt,twoside]{article}

\usepackage{asp2006}
\usepackage{epsf}
\usepackage{psfig}
\usepackage{lscape}
\usepackage{graphicx}
\usepackage{subfigure}

\begin{document}
\title{The NUGA project: The Seyfert 2 galaxy NGC 3147}   
\author{V. Casasola$^{1,2}$, F. Combes$^{2}$, S. Garc\'{\i}a-Burillo$^{3}$, L. K. Hunt$^{4}$,
  S. L\'eon$^{5}$, A. Baker$^{6}$}   
\affil{$^{1}$Dip. di Astronomia Padova, $^{2}$Observatoire de Paris-LERMA, $^{3}$Observatorio Astron\'omico Nacional-OAN Madrid,
$^{4}$Istituto di Radioastronomia-INAF Firenze, $^{5}$Instituto de Astrof\'{\i}sica de Andaluc\'{\i}a CSIC Granada,
$^{6}$Dep. of Physics and Astronomy, Rutgers New Jersey}   

\begin{abstract} 
We present CO(1-0) and CO(2-1) maps of the Seyfert 2 galaxy NGC 3147 
of the NUclei of GAlaxies (NUGA) sample at 1.8$''$ $\times$ 1.6$''$ and 1.4$''$ $\times$ 1.2$''$ 
resolution respectively. Identifying the presence of a bar in NGC 3147
we also compute the gravity torques exerted from the stellar bar on the gas.
\end{abstract}


\section{Molecular gas distribution in NGC 3147:  CO rings}   
\begin{figure}
\centering
\subfigure
{\includegraphics[width=5cm,angle=-90]{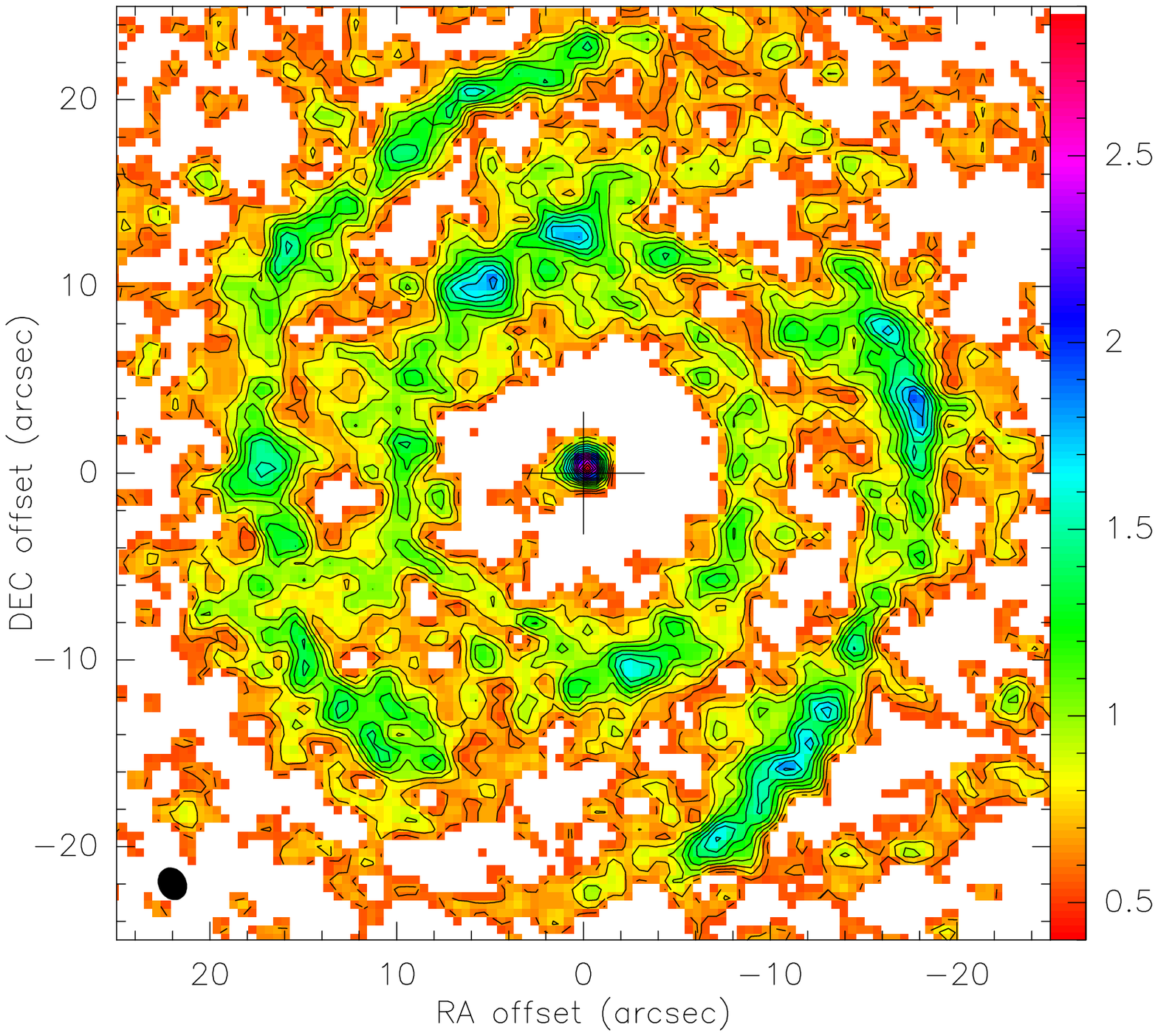}}
\hspace{5mm}
\subfigure
{\includegraphics[width=5cm,angle=-90]{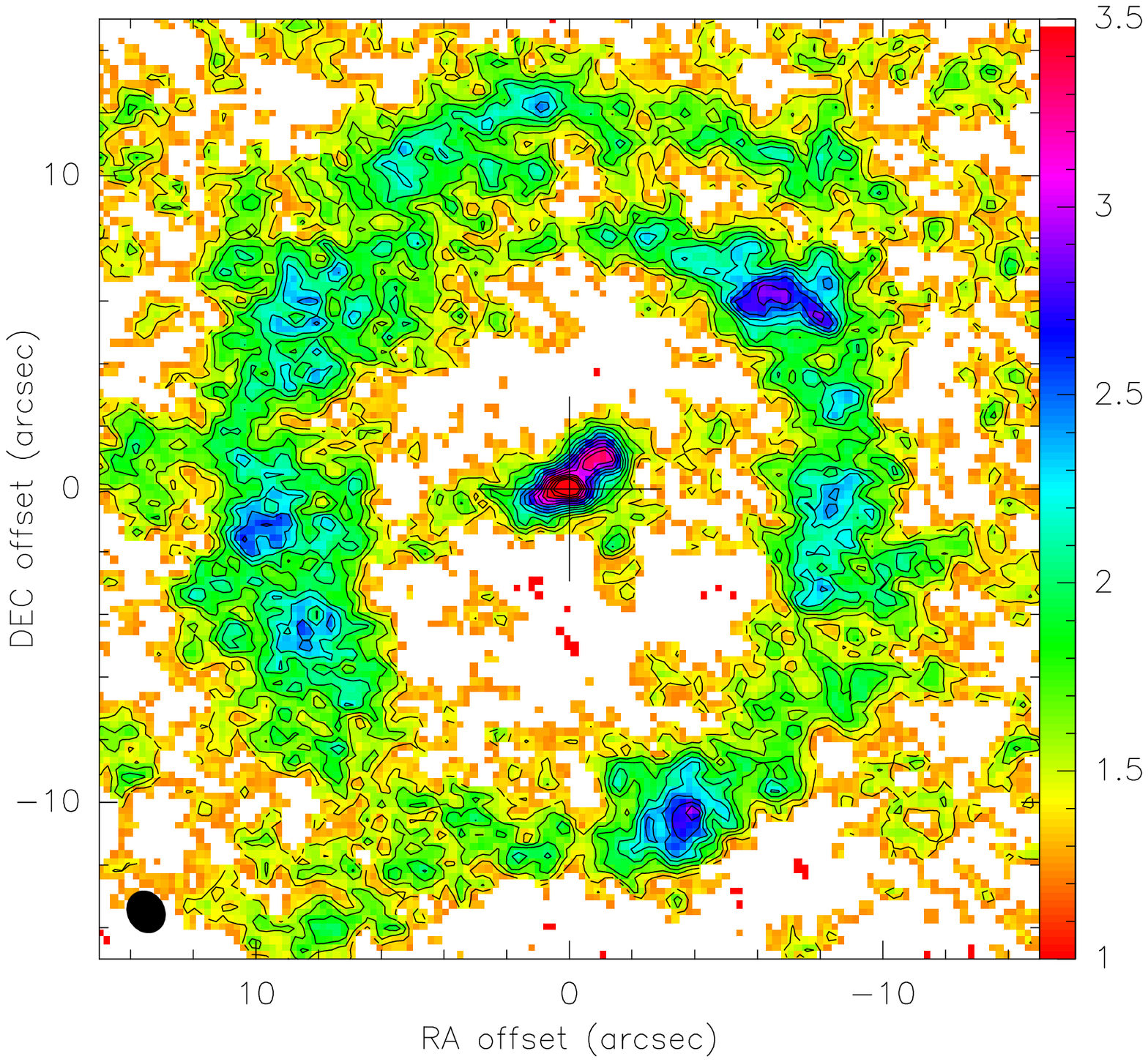}}
\caption{\textit{Left panel:} CO(1-0) integrated intensity contours observed with the IRAM PdBI
	 toward the center of NGC~3147.  
	 Contour levels are from 2$\sigma$ to
	 17$\sigma$ with 1$\sigma$ (= 0.15 Jy/beam km s$^{-1}$) spacing. \textit{Right panel:} 
	 Same as \textit{Left panel} for the CO(2-1) line.
	The rms noise level is
	$\sigma$ = 0.2 Jy/beam km s$^{-1}$.
	Contour levels are from 2$\sigma$ to
	18$\sigma$ with 1$\sigma$ (= 0.2 Jy/beam km s$^{-1}$) spacing.}
	\label{fig1}
\end{figure}

NUGA project \citep{garcia03} is a combined effort to 
undertake a high resolution (1$''$-2$''$) and high-sensitivity 
CO survey, made with the Plateau de Bure IRAM interferometer, 
for a sample of nearby AGN spirals.\\
For the galaxy NGC 3147, after merging the interferometer data with the 
short spacing observations done with the IRAM-30m, we find that a 
nuclear peak and two molecular rings or pseudo-rings 
(at r $\sim$10$^{\prime\prime}$ $\simeq$ 2 kpc  and r $\sim$20$^{\prime\prime}$ $\simeq$ 4 kpc, 
respectively) dominate the CO(1-0) map, while in the CO(2-1) we detect 
a peak of emission in the nucleus and a 
single and central ring at r $\sim$10$^{\prime\prime}$ (Fig. \ref{fig1}).

\section{Comparison between CO and star formation tracers}
\begin{figure}[!h]
\centering
\subfigure
{\includegraphics[width=5cm,angle=-90]{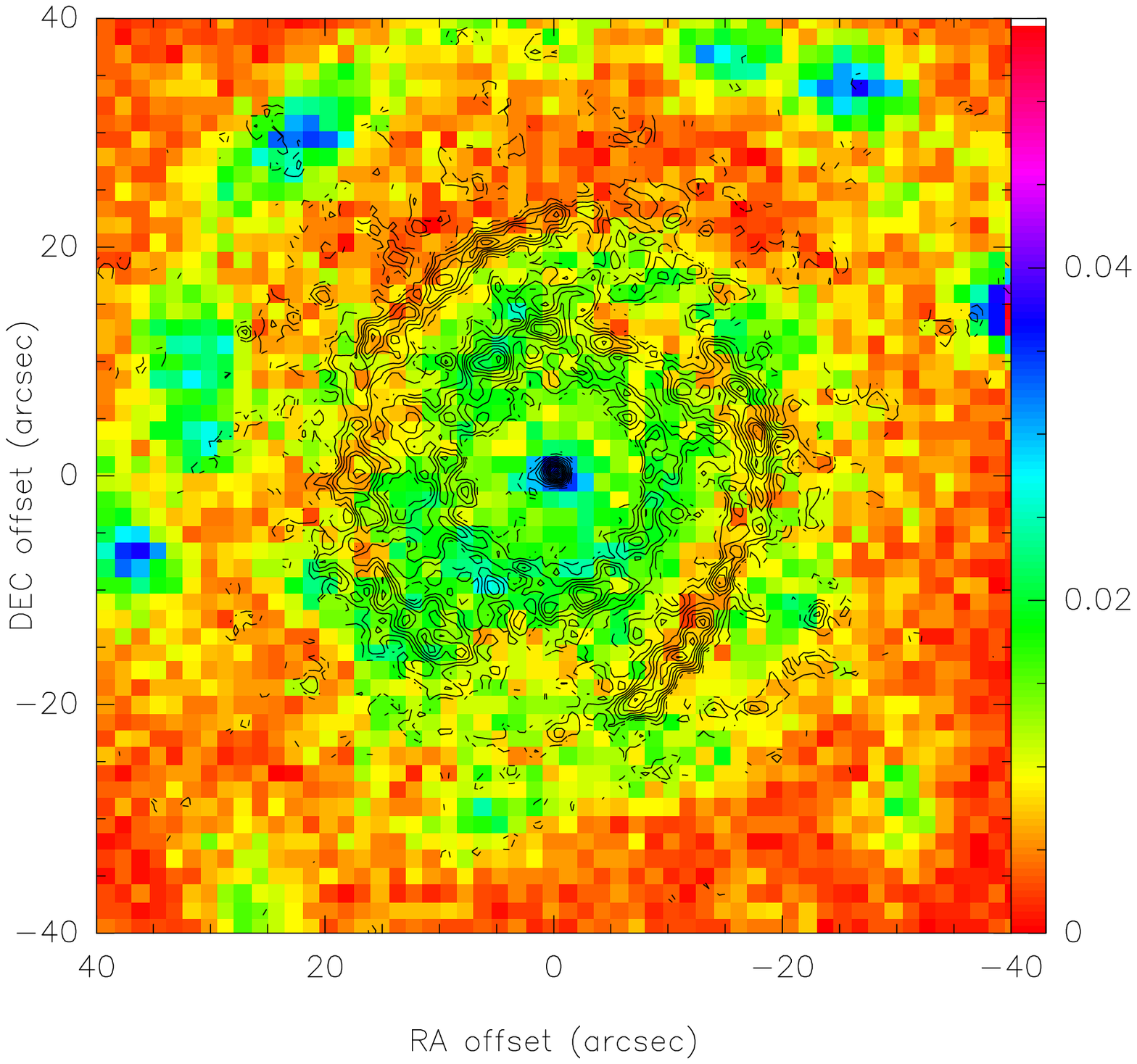}}
\hspace{5mm}
\subfigure
{\includegraphics[width=5cm,angle=-90]{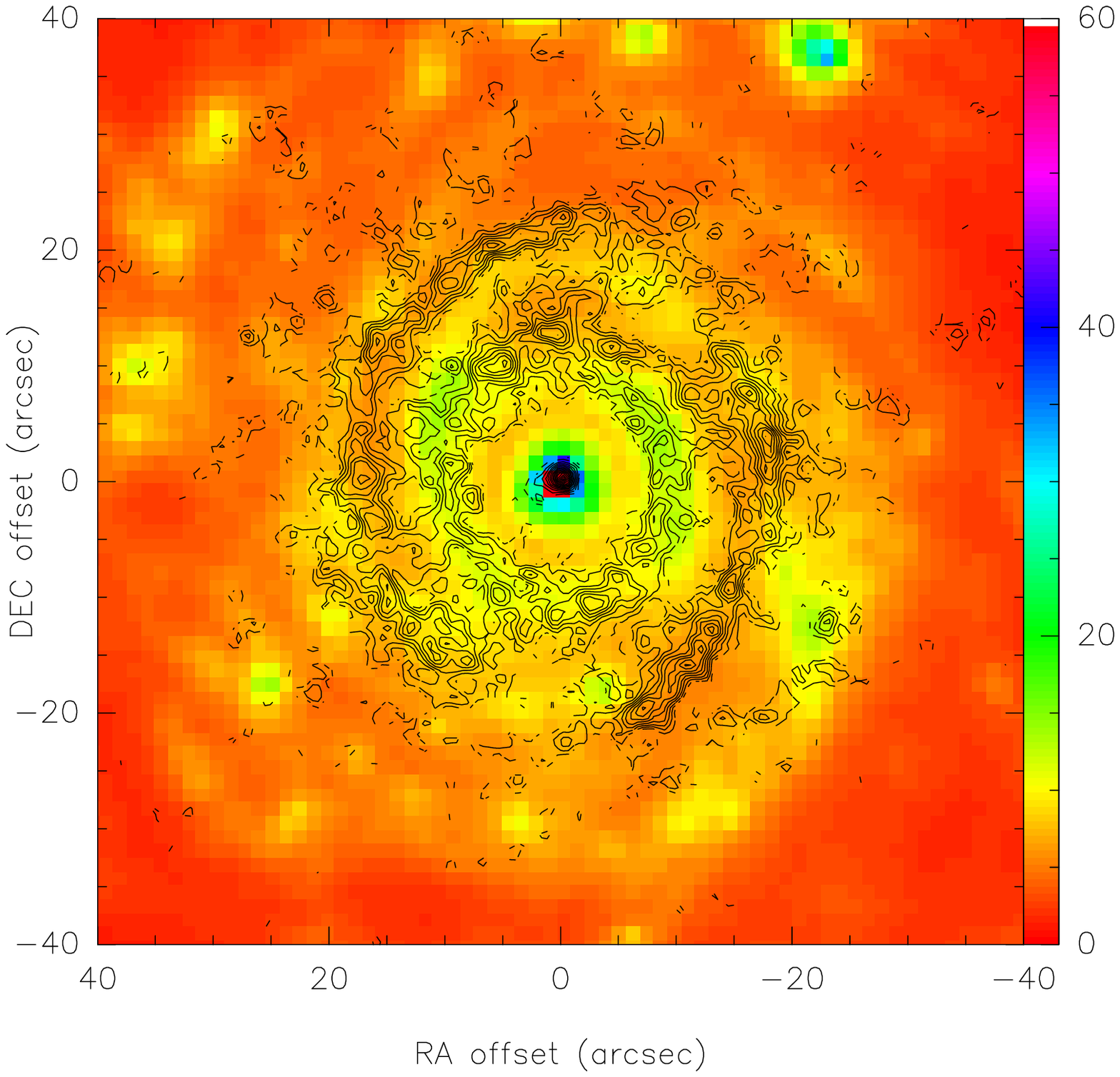}}
\caption{\textit{Left panel:} Superposition of the CO(1-0) contour levels and the FUV-GALEX image. 
	\textit{Right panel:} Superposition of the 
  CO(1-0) contour levels and the 8 $\mu$m SPITZER/IRAC. 
	There is a good agreement between the CO
	and the FUV and mid-IR emissions in the nuclear peak and in the inner ring, but the 
	external CO pseudo-ring 
	is found in the FUV and dust ``interarm'' regions.}
\label{fig2}
\end{figure}

Fig. \ref{fig2} shows that the molecular gas distribution has partial counterparts at
wavelengths tracing the star formation, e.g. infrared emission (SPITZER) 
and ultraviolet one (GALEX). 

\section{Computation of the torques}
Using a NIR image obtained at the 
Canada-France-Hawaii Telescope, we identify the presence of 
a bar in NGC 3147 classified as non-barred galaxy in the optical. 
Computing the gravity torques from 
this stellar bar, we find that the gas in the main central CO pseudo-ring is subject to a 
dominant negative torque and loses angular momentum.  
On the contrary, the gas in spiral arms forming the external 
pseudo-ring suffers positive torques and is driven outwards.
We conclude that some molecular gas is presently inflowing to the center, and probably feeding the low-luminosity
Seyfert 2 observed in NGC 3147 \citep[see][]{casasola07}.


\acknowledgements 
We would like to thank SOC and LOC of the conference for supporting one of us (V. Casasola).


\end{document}